\newcounter{myctr}

\documentclass{myarticleclass}

\usepackage{amssymb,amsfonts,amsmath}
\usepackage{epsfig}
\usepackage{psfrag}

\newcommand{\tp}{^\mathrm{T}} 		
\newcommand{\Prob}{\operatorname{Pr}}	
\newcommand{\Sh}{\operatorname{H}}	
\newcommand{\I}{\operatorname{I}}		

\begin{document}

\catchline{}{}{}{}{}

\title{Inference of higher order substitution dynamics\\ by Markov chain lumping}

\author{\footnotesize OLOF G\"ORNERUP and MARTIN NILSSON JACOBI}

\address{Complex Systems Group, Department of Energy and Environment\\
Chalmers University of Technology, 412 96 G\"{o}teborg, Sweden\\
olofgo@chalmers.se, mjacobi@chalmers.se}

\maketitle

\begin{abstract}

We apply Markov chain lumping techniques to aggregate codons from an empirical substitution matrix. The standard genetic code as well as  higher order amino acid substitution groups are identified. Since the aggregates are derived from first principles they do not rely on system dependent assumptions made beforehand, e.g.~regarding criteria on what should constitute an amino acid group. We therefore argue that the acquired aggregations more accurately capture the multi-level structure of the substitution dynamics than alternative techniques. 

\end{abstract}


\section{Introduction}
\label{sec:sec_intro}

Ever since its discovery by Nirenberg et al.~\cite{Nirenberg61}, the organization of the genetic code has been studied extensively. One main theme of research concerns the organization with respect to physico-chemical properties of the codons, amino acids and amino acid groups. Similar codons are for instance associated with amino acids with similar properties  \cite{Woese65b} and amino acids with simple structures are typically coded by more codons \cite{DiGiulio05}. Many of these studies are based on a snapshot of the code, where static information of what codes to what is used. Although amino acids may be grouped with respect to specific properties, it is difficult to quantitatively judge the relative relevance of these properties. From this standpoint it may prove more conclusive to study the evolutionary dynamics acting on codons and amino acids. The amino acid substitution process is often modeled as a Markov chain, where the distribution of substitutions of a given residue is independent of neighboring residues as well as prior residues at the same site. These assumptions are clearly false, but meaningfully suffices as approximations. Dayhoff and coworkers estimated substitution frequencies empirically from alignments of related sequences \cite{Dayhoff78}. From inspection of log odds scores they concluded that amino acids with similar properties indeed tend to form groups that are conserved: Members of a group substitute with a high frequency internally compared to substitution frequencies to external amino acids. This is natural since replacements within such groups are less likely to have a harmful effect. By using the transition matrix provided by Dayhoff et al.~as a direct estimate of similarity,  French et al.~applied multi-dimensional scaling to demonstrate that amino acids that are likely to mutate into each other more specifically are correlated with respect to side chain volume and hydrophobicity \cite{French83}. These findings are in line with more recent work, e.g.~by Wu et al~\cite{Wu96}, who derived substitution groups from empirical statistics by the same conservation criterion.

In this paper we present an alternative and more general approach to amino acid substitution groups that is based on the concept of Markov chain lumping \cite{Kemeny76}. For this study we use empirical codon substitution frequencies provided by Schneider et al.~\cite{Schneider05}. These were estimated from 17,502 pairwise alignments of orthologous sequences from human, mouse, chicken, frog and zebrafish. 8.3 million codons were aligned and the substitutions between codons were counted. A substitution probability matrix was derived from the resulting counts. For this purpose they utilized the current complete vertebrate genome databases from ENSEMBL \cite{Hubbard05}. By employing two complementary Markov lumping techniques \cite{Meila,Jacobi07}, we infer aggregated representations of Schneider et al.'s substitution probability matrix. These techniques apply to linear dynamical systems in general (including Markov chains) and may therefore in addition be used to analyze hierarchical organization in other types of biological systems. Since the lumping criterion is derived from first principles it is independent of interpretations of the specific system at hand. Assumptions e.g.~regarding amino acid conservation or group isolation, in the specific case of codon substitutions, are therefore not necessary. 

\section{Method}
\label{method}
\begin{figure}
\begin{center}
\includegraphics[scale=0.65]{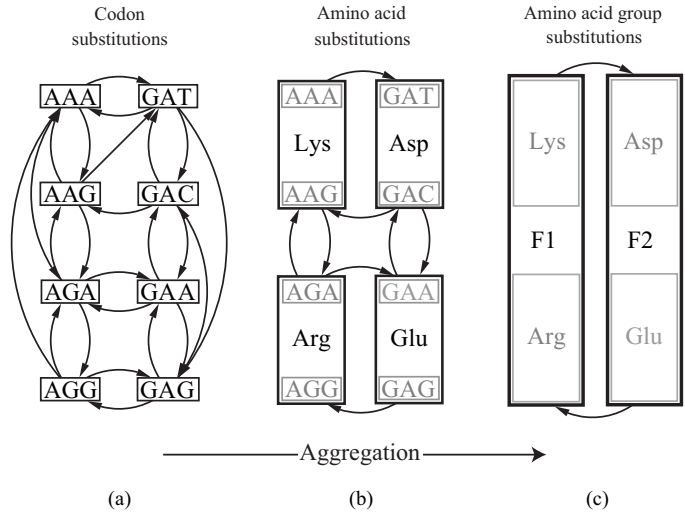} 
\caption{Markov chain lumping; three levels of dynamics. (a) Codon substitutions are modeled as a Markov chain. States represent codons and transitions represent substitutions between codons. (b) If the codons are aggregated with respect to the amino acids they code for, the new aggregated process remains a Markov chain. For instance, since AAA and AAG both code for lysine, they can be aggregated into one unit. (c) Specific aggregates of amino acids also exist such that their dynamics is Markovian. Lysine and arginine can for example be merged to form one state.}
\label{fig:schematic}
\end{center}
\end{figure}

Aggregation of a Markov chain means that the state space is partitioned by an agglomeration of the original states into macro-states. The result is a reduced process with fewer states than the original Markov chain. In general an aggregated dynamics is not a Markov chain as the reduced process has longer memory than the original process. In the special cases when the process over the partitions are a  Markov chain with the same order as the original process, we say that the aggregation is a lumping. An important special case of lumpability is when the dynamics has different time scales. Approximate lumps can then defined as meta-stable states and the Markov transitions are characterized by much more frequent jumps within the lumps than between the lumps. This situation is closely related to the definition of communities in networks, a subject that has been extensively studied recently~\cite{Newman,E}. It turns out that the codon substitution process is approximately lumpable since it can be organized into meta-stable aggregated states on several levels. See Fig.~\ref{fig:schematic} for a schematic illustration of Markov chain lumping of substitution dynamics.

In general, the degree by which a coarser process fulfills the Markov criteria can be measured as the expected mutual information, $\langle \I \rangle$, between the process' past and future states, given its current state. Let $\{s_1, s_2, ..., s_n\}$ be the state space of an aggregated process, $P_i$ a stochastic variable of the past states preceding $s_i$, and $F_i$ a stochastic variable of the subsequent state of $s_i$. The mutual information between past and future states, given a current state $s_i$ is 
\begin{equation}
\I(P_i; F_i)=\Sh(P_i)+\Sh(F_i)-\Sh(P_i, F_i),
\label{MutInfoEq}
\end{equation}
where $\Sh(P_i)$ is the Shannon entropy
\begin{equation}
\Sh(P_i)=- \sum_{j=1}^n \Prob(P_i=s_j) \log_2 \Prob(P_i=s_j)
\end{equation}
of $P_i$. $\Sh(F_i)$ and $\Sh(P_i, F_i)$ of the joint distribution of $P_i$ and $F_i$ are defined analogously. Then
\begin{equation}
\langle \I \rangle = \sum_{i=1}^n \Prob(s_i) \I(P_i; F_i),
\label{ExpIEq}
\end{equation}
where $\Prob(s_i) $ is probability that the system is in state $s_i$. The criterion can be used to test whether or not a given partition defines a lumping, but it is not necessarily useful for constructing the partitions that define lumpings. Since the number of possible ways to partition a state space of $N$ states is astronomical even for relatively small $N$ it is not feasible to evaluate all partitions. However, using a spectral decomposition of the transition matrix it is possible derive lumped Markov chains, even when the number of states is relatively large (on the order of $10^3$ or more if the transition matrix is sparse). This is due to the following observation~\cite{Meila,Jacobi07}: Consider $n$ eigenvectors of the transition matrix. These will define $N$ points in an $n$ dimensional space, where each point is associated with a state in the Markov chain. \emph{If the $N$ points form $n$ clusters, these clusters define an aggregation, where aggregates of states are given by corresponding points within clusters.} See \ref{sec:lump} for details. Spectral lumping tends to work well when the time scale separation in the dynamics is strong, or more generally when lumpings are distinct. For the codon substitution process we will see that the genetic code is clearly detected using the spectral method, whereas aggregations of amino acids cannot be detected by the same technique. In the latter case we instead employ a direct agglomeration method. Starting with each state in a separate lump, the state space is successively aggregated by joining the pair of lumps that gives the best lumping result according to the mutual information criterion in Eq.~\ref{ExpIEq}. The agglomeration method is expected to give good results on the first few levels in the aggregation hierarchy but become more unreliably at the more coarse levels.

\section{Results}

\begin{figure}
\begin{center}
\includegraphics[scale=0.7]{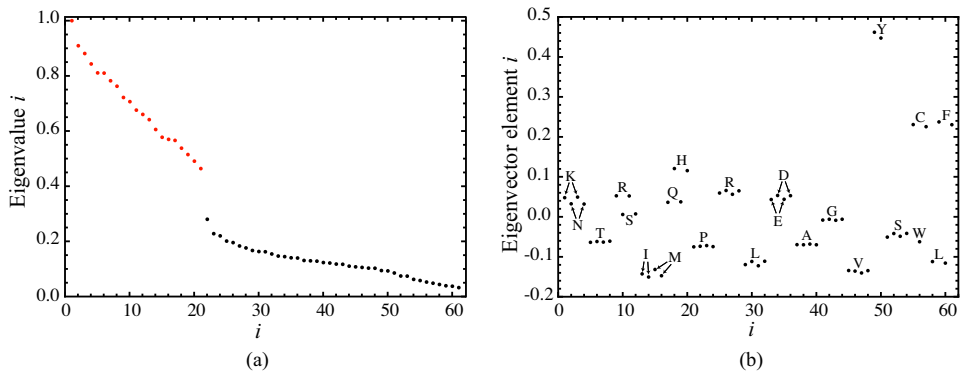} 
\caption{({\bf a}) Eigenvalues of the codon substitution transition matrix $P$. A distinct spectral gap after the $21$ first eigenvalues (marked in red) suggests that the $21$ first eigenvectors reveal an aggregation. ({\bf b}) Vector elements of the fourth eigenvector of $P$ are organized in level sets, where codons that map to the same amino acid are on the same level (with the exception of serine). All the eigenvalues are real because the transition matrix $P$ is reversible.}
\label{fig:spectrum}
\end{center}
\end{figure}

To describe the codon substitution process we use the transition matrix, $P$, provided by Schneider et al. \cite{Schneider05}. Note that the system has $61$ states as substitutions from the three stop codons are not considered. The spectrum of $P$ has a clear gap after the $21$ eigenvalue, Fig.~\ref{fig:spectrum}(a). This gap indicates that the $21$ first eigenvectors may reveal an aggregation of the substitution process~\cite{Meila}. By clustering the elements of the $21$ first eigenvectors of $P$--- resulting in 61 points in a $21$ dimensional space---$21$ distinct clusters are acquired. Since the number of eigenvectors equals the number of clusters, these define a lumping. As exemplified in Fig.~\ref{fig:spectrum}(b) the clusters show as level sets in the individual eigenvectors. Each cluster constitutes codons that are associated with the same amino acid, with the exception of the codons of serine, which are divided into two clusters. This unique separation is due to that serine is the only amino acid whose codons are not connected with single point mutations (i.e.~some codons are separated by a Hamming distance larger than one on a hypercube). The aggregation of individual codons to amino acids, which we know as the standard genetic code, reflects that the functionality of member codons within an aggregate are invariant under mutations as they code for the same amino acid.

\begin{figure}
\begin{center}
\includegraphics[scale=1.0]{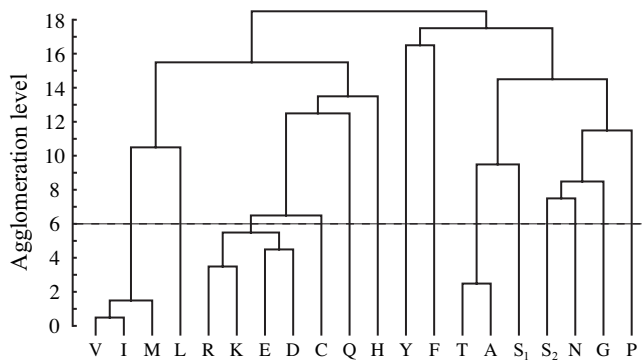} 
\caption{A dendrogram of the result of an agglomeration based on successively joining pairs of states or lumps that result in the best lumping with respect to the mutual information measure in Eq.~\ref{ExpIEq}. The dashed line marks the most significant aggregation, which is also shown in Fig.~\ref{fig:best_aggr}. S$_1$ denotes serine coded by TCT, TCC, TCA and TCG, and S$_2$ denotes serine coded by AGT and AGC.}
\label{fig:agg_dendr}
\end{center}
\end{figure}

\begin{figure}
\begin{center}
\includegraphics[scale=0.8]{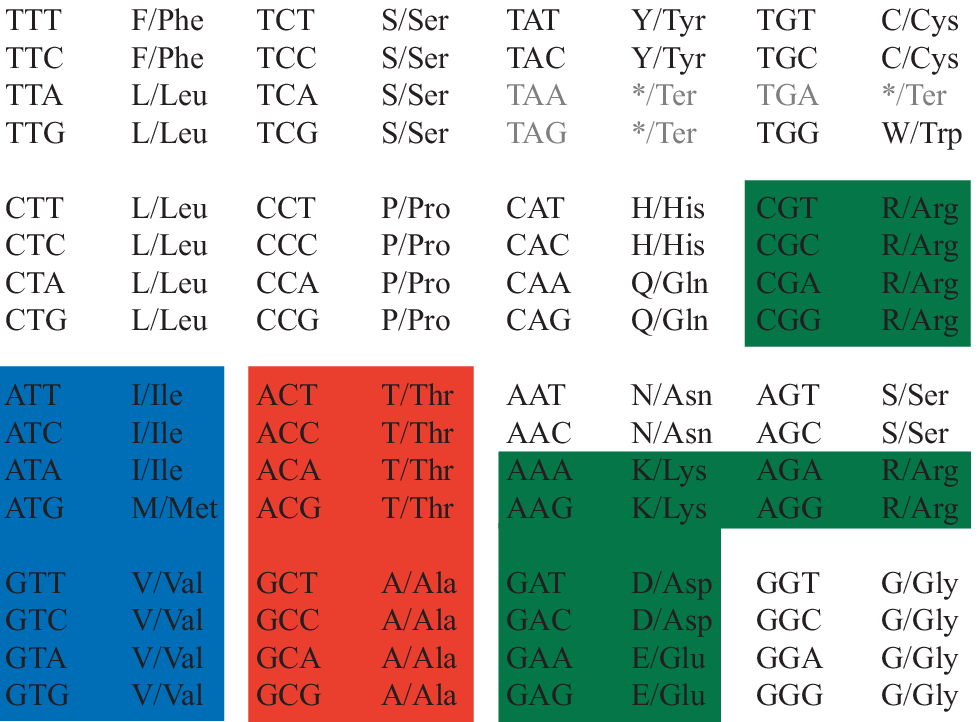} 
\caption{Amino acid groups resulting in the most significant lumping $ \{A, T\}$, $\{I, M, V\}$ and $\{E, D\}$ as shown in the standard genetic code table.} 
\label{fig:best_aggr}
\end{center}
\end{figure}

At the aggregated level of amino acid substitution,  lumpings are not as clearly revealed by the eigenvectors. This is to be expected since the redundancy in the genetic code reflects a much stronger neutrality than possible similarities between the amino acids. If the partitioning of the state space is viewed as an optimization problem aiming to minimize the mutual information defined in Eq.~\ref{ExpIEq}, then there are many almost equivalent minima. In this situation the spectral method does not function well. It is nevertheless possible to identify significant lumpings of amino acids into families that tend to be conserved by the substitution process. Two generic optimization methods are used: Simulated annealing and the agglomeration method described in Section~\ref{method}.  Due to that tryptophan (W) has very low mutability and is the least occurring amino acid, a significant two-state lumping exists where tryptophan forms one aggregate and the rest of the amino acids form another aggregate. To simplify further analysis thryptophan is therefore discarded. The result from the agglomerative aggregation is shown in Fig.~\ref{fig:agg_dendr}. The most significant lumping found using either method has the aggregates $ \{A, T\}$, $\{I, M, V\}$ and $\{E, D\}$ and is shown in Fig.~\ref{fig:best_aggr}.

\section{Discussion}

Markov lumping based on empirical substitution frequencies in coding DNA sequences provides a natural rationale for grouping codons and amino acids. Due to its construction the lumpability condition has a clearer connection to the evolutionary dynamics than previous aggregation methods suggested in the literature, such as multi-dimensional scaling of the transition matrix~\cite{French83} or the compactness and isolation criteria used in \cite{Wu96}. Under the assumption that the codon substitution process can be approximately  described as a Markov process, we argue that  the lumping is an effective method for analyzing the structure of the genetic code and its higher level organization into approximately conserved amino acid families. 

One may hypothesize that aggregated dynamics of codon substitutions provide information about the origin of the genetic code. There are several theories aiming  to address the fundamental question on how the code came to be. See Ref.~\cite{Trifonov00} for a comprehensive comparison. With the exception of the \emph{frozen accident} theory by Crick \cite{Crick68}, these theories couple the evolution of the genetic code primarily with physico-chemical properties of the amino acids or evolved biosynthetic pathways. Woese \cite{Woese65}, specifically, suggested that the code has evolved by a process of ambiguity reduction. The idea is that a crude primordial version of the code, where groups of codons code for groups of amino acids with resembling properties, evolved into the code's current state by a series of refinements. May amino acid groups reflect earlier versions of the code?  Riddle et al.~\cite{Riddle97} experimentally searched for a minimum set of amino acids capable of forming complex protein folds. They found that the five amino acids A, G, I, E and K are capable of forming most of the ancient SH3 protein domain. Consider the aggregation in Fig.~\ref{fig:best_aggr}. A and I are members of separate aggregates, G form its own aggregate; and E and K are in separate aggregates prior to the last agglomeration leading to the aggregation, Fig.~\ref{fig:agg_dendr}. One could speculate that some of these aggregates reflect group codons in an earlier version of the code, and that these groups were specialized into present day codons. Such a view is partly supported Jim\'enez-Montaño's hypothesis on the evolutionary history of the code, which is based on group theory and the thermodynamics of codon-anticodon interactions  \cite{Jimenez99}. In the proposed evolutionary tree, amino acids within aggregates $\{A, T\}$, $\{I, M, V\}$ and $\{E, D\}$ share the same branches up till the two last reassignment of codons. However, a careful analysis would be required in order to conclusively relate acquired aggregates to the evolution of the standard genetic code and its deviates.\\

\noindent {\bf Acknowledgments}: This work was funded in part by the EU integrated project FP6-IST-FET PACE, by  EU FP6-NEST project EMBIO, and by EU STREP project FP6-IST-STREP MORPHEX. The authors thank Rickard Sandberg for helpful discussions.

\bibliographystyle{unsrt}
\bibliography{codon_aggregates}

\begin{thebibliography}{10}

\bibitem{Nirenberg61}
M.~W. Nirenberg and J.~H. Matthaei.
\newblock The dependence of cell-free protein synthesis in e. coli upon
  naturally occuring or synthetic polyribonucleotides.
\newblock {\em Proceedings of the National Academy of Sciences}, 47:1588--1602,
  1961.

\bibitem{Woese65b}
C.~R. Woese.
\newblock Order in the genetic code.
\newblock {\em Proceedings of the National Academy of Sciences}, 54:71--75,
  1965.

\bibitem{DiGiulio05}
M.~Di~Giulio.
\newblock The origin of the genetic code: theories and their relationships, a
  review.
\newblock {\em {BioSystems}}, 80:175--184, 2005.

\bibitem{Dayhoff78}
M.~O. Dayhoff, R.~M. Schwartz, and B.~C. Orcutt.
\newblock {A model of evolutionary change in proteins}.
\newblock In {\em Atlas of protein sequence and structure}, volume~5, pages
  345--358. M. O. Dayhoff, National biomedical research foundation, Washington
  DC., 1978.

\bibitem{French83}
S.~French and B.~Robson.
\newblock What is a conservative substitution?
\newblock {\em Journal of Molecular Evolution}, 19(2), 1983.

\bibitem{Wu96}
Thomas~D. Wu and Douglas~L. Brutlag.
\newblock Discovering empirically conserved amino acid substitution groups in
  databases of protein families.
\newblock In {\em Proceedings of the Fourth International Conference on
  Intelligent Systems for Molecular Biology}, pages 230--240. AAAI Press, 1996.

\bibitem{Kemeny76}
J.~G. Kemeny and J.~L. Snell.
\newblock {\em Finite {M}arkov Chains}.
\newblock Springer, New York, NY, USA, 2nd edition, 1976.

\bibitem{Schneider05}
A.~Schneider, G.~M. Cannarozzi, and G.~H. Gonnet.
\newblock Empirical codon substitution matrix.
\newblock {\em BMC Bioinformatics}, 6(13), 2005.

\bibitem{Hubbard05}
T.~Hubbard, D.~Andrews, M.~Caccamo, G.~Cameron, Y.~Chen, M.~Clamp, L.~Clarke,
  G.~Coates, T.~Cox, F.~Cunningham, V.~Curwen, T.~Cutts, T.~Down, R.~Durbin,
  X.~M. Fernandez-Suarez, J.~Gilbert, M.~Hammond, J.~Herrero, H.~Hotz, K.~Howe,
  V.~Iyer, K.~Jekosch, A.~Kahari, A.~Kasprzyk, D.~Keefe, S.~Keenan,
  F.~Kokocinsci, D.~London, I.~Longden, G.~McVicker, C.~Melsopp, P.~Meidl,
  S.~Potter, G.~Proctor, M.~Rae, D.~Rios, M.~Schuster, S.~Searle, J.~Severin,
  G.~Slater, D.~Smedley, J.~Smith, W.~Spooner, A.~Stabenau, J.~Stalker,
  R.~Storey, S.~Trevanion, A.~Ureta-Vidal, J.~Vogel, S.~White, C.~Woodwark, and
  E.~Birney.
\newblock Ensembl 2005.
\newblock {\em Nucleic Acids Res}, 33(Database issue), January 2005.

\bibitem{Meila}
M.~{Meil\u{a}} and J.~Shi.
\newblock A random walks view of spectral segmentation.
\newblock In {\em In AI and Statistics (AISTATS)}, 2001.

\bibitem{Jacobi07}
M.~Nilsson~Jacobi and O.~G\"{o}rnerup.
\newblock A dual eigenvector condition for strong lumpability of markov chains,
  2008.
\newblock Submitted. arXiv:0710.1986v2.

\bibitem{Newman}
M.~Newman.
\newblock Modularity and community structure in networks.
\newblock {\em Proceedings of National Academy of Sceinces},
  103(23):8577--8582, 2006.

\bibitem{E}
W.~E, T.~Li, and E.~Vanden-Eijnden.
\newblock Optimal partition and effective dynamics of complex networks.
\newblock {\em Proceedings of National Academy of Sciences}, February 2008.

\bibitem{Trifonov00}
E.~N. Trifonov.
\newblock Consensus temporal order of amino acids and evolution of the triplet
  code.
\newblock {\em Gene}, 261:139--151, Dec 2000.

\bibitem{Crick68}
F~H Crick.
\newblock The origin of the genetic code.
\newblock {\em Journal of Molecular Biology}, 38(3):367--379, 1968.

\bibitem{Woese65}
C.~R. Woese.
\newblock On the evolution of the genetic code.
\newblock {\em Proceedings of the National Academy of Sciences}, 54:1546--1552,
  Dec 1965.

\bibitem{Riddle97}
D.~S. Riddle, J.~V. Santiago, S.~T. Bray-Hall, N.~Doshi, V.~P. Grantcharova,
  Q.~Yi, and D.~Baker.
\newblock Functional rapidly folding proteins from simplified amino acid
  sequences.
\newblock {\em Nature Structural and molecular biology}, 4:805--809, Oct 1997.

\bibitem{Jimenez99}
M.~A. {Jim\'enez-Montaño}.
\newblock Protein evolution drives the evolution of the genetic code and vice
  versa.
\newblock {\em {BioSystems}}, 54:47--64, 1999.

\end{thebibliography}

\appendix

\section{Spectral lumping method}
\label{sec:lump}

Consider a Markov chain with $N$ states and a probability distribution vector at time $t$ denoted as $x (t) $. The $N \times N$ transition matrix $P = [ p_{ij} ]$ defines the evolution of the probability distribution $x (t+1) = x(t) P$. A partition of the state space can be defined by a $N \times K$ matrix $\Pi = [ \pi _{ik} ]$, where $K$ is the number of aggregates and $\pi _{ik} =1$ if state $i$ is in aggregate $k$ and $0$ otherwise. The matrix $\Pi$ consists of only $0$'s and $1$'s with exactly one $1$ in each row. A necessary and sufficient criterion for a partitioning to define a lumping of a Markov chain is that for each state in a specific aggregate the total transition probability into other aggregates are equal. For a transition matrix $P$ this can be expressed as $\sum _j P _{ij} \Pi _{jk}$ being constant for all $i$ within an aggregates and $k$ \cite{Kemeny76}. Or equivalently, there must exist a $K \times K$ matrix $Q$ such that

\begin{equation}
	P \Pi = \Pi Q ,
	\label{eq:reduction}
\end{equation}
in which case the reduced probability distribution is defined by $y (t) = x (t) \Pi$ and $Q = [ q _{ij} ]$ is the transition matrix for the lumped process. We note that the left hand side of Eq.~\ref{eq:reduction} produces a new matrix by $P$ acting on each column in $\Pi$, whereas on the right hand side the matrix is constructed by each column being a linear combination of the columns in $\Pi$ with linear coefficients $q_{ij}$. This implies that there exist a matrix $Q$ satisfying Eq.~\ref{eq:reduction}  if and only if the columns of $\Pi$ span a left invariant subspace of $P$. If $P$ is diagonalizable, which is the case for the codon substitution process, then any (right) invariant subspace can be spanned by the (right) eigenvectors.

If the columns of $\Pi$ are chosen to be a subset of the eigenvectors, then $Q$ is just a diagonal matrix with the corresponding eigenvalues on the diagonal. This choice does define a reduction of a linear system but not usually a lumping of a Markov chain. The columns of $\Pi$  must contain exactly one $1$ and the rest of the elements must be $0$. At the same time the $K$ columns must be linear combinations of a set of $K$ eigenvectors of $P$. To see how these two requirements can be met, consider an example of a lumping matrix that aggregates four states into two. The two columns of the $\Pi$ matrix are, $\left( 1 , 0 , 0 , 1 \right) \tp$ and $\left( 0, 1,1,0 \right) \tp$. Any linear combination of the columns has the structure $( a , b , b, a ) \tp$, i.e. the elements that are lumped into the same aggregate are identical. From this observation it is possible to show that a necessary and sufficient condition for lumping a Markov chain $x(t+1) = x(t) P$ into $K$ lumps is that there exist $K$ right eigenvectors of $P$ whose elements constant within the aggregates \cite{Jacobi07}. 

Given a set of right eigenvectors, $v_{i_1} , v_{i_2} , \dots , v_{i_K}$, a suggested lumping can be identified using a clustering algorithm, e.g. K-mean. For each of the $N$ states in the Markov chain the elements of the eigenvectors define a $K$ dimensional coordinate. States that are in the same aggregate should have elements that are close together in all eigenvectors, e.g.~as in Fig.~\ref{fig:spectrum}(b), and therefore form clusters in the $K$-dimensional space. 

In practice, the lumpability condition (\ref{eq:reduction}) is never fulfilled exactly. It is therefore important to evaluate the quality of an approximate lumping. It is of course possible to use Eq.~\ref{eq:reduction} directly, for example by defining the quality of a lumping as e.g. the norm $\| P \Pi - \Pi Q \| _2$, which should be minimized. However, the fundamental idea behind Markov chain lumping is to find aggregated Markov dynamics whose order is not increased, i.e. does not have longer memory than the original process. Since memory is naturally defined in terms of the mutual information between the past and the future conditioned on the present, as in Eq.~\ref{ExpIEq}, we use this as a lumping measure.

\end{document}